\documentstyle[prl,aps,multicol,epsfig]{revtex}
\begin{document}

\title{A model for correlations in stock markets}
\author{Jae Dong Noh}
\address{Center for Theoretical Physics, Seoul National University,
Seoul 151-742, Korea}

\maketitle

\begin{abstract}
We propose a group model for correlations in stock markets.
In the group model the markets are composed of several groups,  
within which the stock price fluctuations are correlated. 
The spectral properties of empirical correlation matrices reported in
[Phys. Rev. Lett. {\bf 83}, 1467 (1999); Phys. Rev. Lett. {\bf 83}, 
1471 (1999.)] are well understood from the model. It provides the connection 
between the spectral properties of the empirical correlation matrix and 
the structure of correlations in stock markets.
\end{abstract}
\draft
\pacs{PACS numbers: 87.23.Ge, 02.10.Sp, 05.40.Ca, 05.45.Tp}

\begin{multicols}{2}

There has been growing interest of physicists in economic systems.  
They exhibit various interesting complex behaviors, e.g., a power-law
distribution of the variations of financial market 
indices~\cite{Mantegna&Stanley,Gopikrishnan,Plerou_preprint}, and  
many efforts have been devoted to understanding
the temporal correlations~\cite{Bak,Lux&Marchesi,Eguiluz_DHulst}. 
The economic systems are composed of a large number of interacting units. 
The information of the correlations and interactions is provided by
a correlation matrix~\cite{Laloux,Plerou,Mantegna}.
Its matrix elements are given by the overlap between price changes of
different assets. They measure a extent to which the price
fluctuations of different assets are correlated. 

Recently Laloux {\em et al.}~\cite{Laloux} and 
Plerou {\em et al.}~\cite{Plerou} studied spectral properties of
the empirical correlation matrix ${\bf C}$ of stock price changes 
in real markets. Comparing the empirical data with 
the theoretical prediction of the random matrix theory, 
they observed a correlated aspect of the financial markets.
However, the nature of the correlations is not revealed yet. 
In this work, we propose that the observed spectral properties 
reflect a specific structure of correlations in the financial markets. 
To this end, we postulate a model 
for the correlations and show that it explains all aspects of the observed 
spectral properties.

Following the notation of~\cite{Laloux}, the empirical correlation matrix is 
constructed from the time series of stock price changes $\delta x_i(t)$, 
where $i$~($i=1,\ldots,N$) labels the assets and $1\leq t \leq T$ the time. 
It is assumed that $\delta x$'s are shifted and rescaled to have zero mean
and a unit variance~\cite{Laloux}. The correlation matrix has elements
\begin{equation}
C_{ij} = (\vec{\delta x}_i) \cdot (\vec{\delta x}_j) \ .
\end{equation}
Here we introduced a symbolic notation that $\vec{x} \cdot \vec{y} =
\frac{1}{T}\sum_{t=1}^T x(t) y(t)$.

Suppose that there exist only random correlations: 
$\delta x_i(t)$ are independent identically-distributed random variables.
In that case the distribution of eigenvalues $\lambda$ of ${\bf C}$
is exactly known in the limit $N,M\rightarrow \infty$ with $Q=T/N\geq 1$
fixed~\cite{Sengupta}.
The density of the eigenvalues is given by
\begin{equation}\label{RMT}
\rho(\lambda) = \frac{Q}{2\pi}
\frac{\sqrt{(\lambda_{\max}-\lambda)(\lambda - \lambda_{\min})}}{\lambda} \ ,
\end{equation}
for $\lambda\in[\lambda_{\min},\lambda_{\max}]$ and zero otherwise, where
$\lambda_{\max} = (1+1/\sqrt{Q})^2$ and $\lambda_{\min} = (1-1/\sqrt{Q})^2$. 
Deviations from Eq.~(\ref{RMT}) provide clues about the structure of the
financial markets and the nature of the underlying interactions.

Here we briefly summarize the finding of Ref.~\cite{Laloux,Plerou}. They
constructed the correlation matrix of $N$ assets from the major stock 
markets and calculated the density of eigenvalues, $\rho(\lambda)$, 
and the inverse participation ratio, $I$. 
It is related to the reciprocal of the number of eigenvector components 
significantly different from zero, and defined as
\begin{equation}
I_k = \sum_{l=1}^N [u_{kl}]^4 \ ,
\end{equation}
where $k$~($k=1,\ldots,N$) refers to the eigenstate and $u_{kl}$ is the
$l$th components of its normalized eigenvector. 
It is observed that $\rho(\lambda)$
follows the distribution in Eq.~(\ref{RMT}) for small
$\lambda\lesssim 2.0$~\cite{Laloux,Plerou}. Those low-lying
eigenstates reflect a random noisy part of the correlation.
Besides, it is also observed that there appear several isolated 
eigenstates for large $\lambda>2.0$~\cite{Laloux,Plerou} and that  
the eigenstates at both ends of the spectrum have large value of
$I$~\cite{Plerou}. However, an appropriate explanation for these 
is not presented.

To explains the observed spectral properties, we propose a model for 
the correlations,  which will be called the group model. 
In the group model it is assumed that
the stock price changes of companies in the same kind of industry have 
a strong correlation. It is in agreement with our everyday experiences and 
a quantitative study supports this~\cite{Mantegna}.
Those sets of companies will be called the groups. The group feature is 
incorporated into the model by assuming that the stock price change 
$\delta x_i(t)$ of a company $i~(=1,\ldots,N)$ at time $t~(=1,\ldots,T)$ 
is given by a sum of two random variables:
\begin{equation}\label{time_series}
\delta x_i (t) = f_{\alpha_i}(t) + \epsilon_i(t) \ ,
\end{equation}
where $\alpha_i$ denotes the group to which a company $i$ belongs, 
$f_{\alpha_i}$ represents a synchronous variation of the group $\alpha_i$, and 
$\epsilon_i$ a random noisy part.  The noise strength is parametrized by 
$\varepsilon_{\alpha_i}$ with which the variances of $f_{\alpha_i}$ and
$\epsilon_i$ are given by $(1+\varepsilon_{\alpha_i}^2)^{-1}$ and
${\varepsilon_{\alpha_i}^2}(1+\varepsilon_{\alpha_i}^2)^{-1}$,
respectively. The correlation matrix takes the form of
${\bf C} = {\bf C}^0 + {\bf C}^1$ with $C^0_{ij} =
\delta_{\alpha_i,\alpha_j}$~($\delta_{\alpha,\beta}$ : the Kronecker 
delta symbol). 

When $\varepsilon_\alpha=0$ and $T$ is infinite,  
${\bf C}={\bf C}^0$ becomes a block diagonal matrix; a direct
product of $N_\alpha\!\times\! N_\alpha$ matrices with all elements being
unity~($N_\alpha$: the size of the group $\alpha$).
Each block has a nondegenerate eigenstate with the eigenvalue 
$\lambda=N_\alpha$ and $(N_\alpha\!-\!1)$-fold degenerate 
eigenstates with $\lambda=0$. 
The inverse participation ratio of the nondegenerate eigenstates is 
the inverse of the group size, $I = 1/N_\alpha$.

For nonzero $\varepsilon_\alpha$ and finite $T$, the random overlap part
${\bf C}^1$ becomes nonzero. Its effect can be treated as a perturbation to 
${\bf C}^0$. 
Up to the first order perturbation, the eigenvalues of the nondegenerate 
states are scaled down by a factor 
$(1+\varepsilon_\alpha^2)^{-1}$ and the degenerate states split 
into the states whose eigenspectrum are given by those 
of ${\bf C}^1$. Therefore, the perturbed degenerate eigenstates with small
$\lambda$ will result in the distribution of the eigenvalues of 
Eq.~(\ref{RMT}), while the nondegenerate states isolated peaks 
at $\lambda\simeq (1+\varepsilon_\alpha^2)^{-1}N_\alpha$ as found
in~\cite{Laloux,Plerou}. 

The information of the correlation is mainly contained in the isolated 
eigenstates. Each one represents a correlated group, whose size and
participating companies are obtained from the eigenvector and its 
eigenvalue.  The eigenstates with small eigenvalues, which are 
approximately the eigenstates of ${\bf C}^1$, may also provide 
the correlation information. 
Since ${\bf C}^1$ generates random overlaps between groups, 
its eigenvector components spread over the whole market in general. 
However, the overlap may be confined mainly within the group, provided that  
the noise strength of a group is small. Then there may appear an eigenstate 
with the eigenvalue near the unperturbed value, zero, and with relatively 
large inverse participation ratio, as observed in~\cite{Plerou}. 

\begin{figure}
\centerline{\epsfig{file=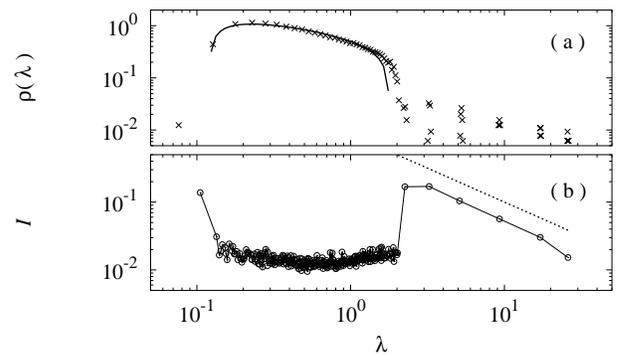,width=86mm}}
\caption{(a) $\rho(\lambda)$~($\times$) and a fitting 
curve to the form in Eq.~(\ref{RMT})~(solid line).
(b) Inverse participation ratio with a plot of $1/\lambda$ against
$\lambda$~(dotted line).}\label{fig1}
\end{figure}

We demonstrate the validity of the group model by examining the spectral 
properties of an artificial correlation matrix constructed from a time
series of Eq.~(\ref{time_series}) with $N=254$ and $T=800$.
A part of them is separated arbitrarily into groups of size 
$N_\alpha=2,4,8,16,32,64$ and the remaining ones are assumed to behave 
independently. 
The noise strength is assigned to $\varepsilon_\alpha= (1-1/N_\alpha)$. 
Figure~\ref{fig1} shows $\rho(\lambda$), ensemble-averaged over 50
realizations of the time series, in (a) and the inverse participation ratio 
$I$ for a realization in (b). 
They are plotted in the log-log scale to show the large
and small $\lambda$ regions on an equal footing.
They are in excellent agreement with the empirical data shown
in~\cite{Laloux,Plerou}. $\rho(\lambda)$ for $\lambda<2.0$
is well fitted to Eq.~(\ref{RMT}). It deviates from Eq.~(\ref{RMT}) for
$\lambda>2.0$. The isolated peaks represents the groups of size
$N_\alpha=4,\ldots,64$. A peak for the group of $N_\alpha=2$ is smeared
with the continuous distribution. The inverse participation ratio is large
at both ends of the spectrum. For large $\lambda$, $\lambda$
and $I$ are inversely proportional to each other since $\lambda$ is
proportional to $N_\alpha$, and $I$ to $1/N_\alpha$. The eigenstate
with the smallest eigenvalue has also large value of $I$. It has
large components within the group of size $N_\alpha=2$, which has smallest
noise strength.

In summary, we proposed the group model to explain the spectral
properties of the empirical correlation matrices. 
According to the group model, the stock markets are composed of several 
correlated groups of companies. The stock price fluctuations of companies 
within a group have a synchronous component. The groups manifest themselves
as the eigenstates of ${\bf C}$ with large eigenvalues.
The sizes of groups and composing companies are known from the eigenvalues
and the eigenvectors. It will be an interesting work to reveal the group
structure of the real markets from the empirical correlations matrix.

The author thanks D. Kim, G. S. Jeon, and H.-W. Lee for helpful discussions. 
This work is supported by the Korea Science and Engineering Foundation 
through the SRC program at SNU-CTP.

\end{multicols}

\end{document}